\documentclass{sig-alternate-05-2015}
\usepackage{rotating} 
\usepackage{multirow}

\usepackage[english]{babel}
\usepackage{blindtext}

\usepackage{etoolbox}
\makeatletter
\patchcmd{\maketitle}{\@copyrightspace}{}{}{}
\makeatother

\begin{document}


\doi{10.475/123_4}

\isbn{123-4567-24-567/08/06}

\conferenceinfo{PLDI '13}{June 16--19, 2013, Seattle, WA, USA}

\acmPrice{\$15.00}

%

\title{Toward  Active Learning   in Cross-domain Recommender Systems}
%
%
%
%
%

\numberofauthors{4} 
%
\author{
%
%
\alignauthor
	 Roberto Pagano \\
       \affaddr{Politecnico di Milano}\\
       \affaddr{Via Ponzio 34/5}\\
       \affaddr{20133, Milano, Italy}\\
       \email{roberto.pagano@polimi.it}
\alignauthor
	 Massimo Quadrana \\
       \affaddr{Politecnico di Milano}\\
       \affaddr{Via Ponzio 34/5}\\
       \affaddr{20133, Milano, Italy}\\
       \email{massimo.quadrana@polimi.it}
      \alignauthor  Mehdi Elahi \\
       \affaddr{Free University of Bozen - Bolzano}\\
       \affaddr{Piazza Domenicani 3}\\
       \affaddr{39100, Bolzano, Italy}\\
       \email{meelahi@unibz.it}
\and  
\alignauthor  Paolo Cremonesi\\
       \affaddr{Politecnico di Milano}\\
       \affaddr{Via Ponzio 34/5}\\
       \affaddr{20133, Milano, Italy}\\
       \email{paolo.cremonesi@polimi.it}
}

\date{30 July 1999}

\maketitle
\begin{abstract}
One of the main challenges in Recommender Systems (RSs) is the {\it New User} problem which happens when the system has to generate personalized recommendations for a new user whom the system has no information about. {\it Active Learning  } tries to solve this problem by acquiring user preference data  with the maximum quality, and with the minimum acquisition cost.

Although there are variety of works in   active learning for RSs research area,  almost all of them have focused only on the single-domain recommendation scenario. However, several real-world RSs 
operate in the cross-domain scenario, where the system generates recommendations in the target domain by exploiting user preferences in both the target and auxiliary domains. In such a scenario, the performance of active learning strategies can be significantly  influenced
and typical active learning strategies may fail   to perform properly.  

In this  paper, we address this limitation, by evaluating active learning strategies in a novel evaluation framework, explicitly suited for the cross-domain recommendation scenario. 
We show that having access to the preferences of 
the users in the auxiliary domain  may have a huge impact on the performance of active learning strategies w.r.t. the classical, single-domain scenario.
\end{abstract}

\section{Introduction}

In general terms, there are two tasks that are mainly performed by Recommender Systems: {\it Learning} the users' preferences, and {\it Recommending} the items to users based on these preferences \cite{rubens2015active}. When a new user registers to a RS, the system has no preference of that user and hence is not able to produce relevant recommendations and this problem still represents a big challenge for RSs \cite{elahi14active}.

 Active Learning attempts to solve this problem by eliciting
preferences of the users and learning   their tastes.
 It does not solely focus on the quantity of the data elicited from the users, but also on the quality of the data.
 Hence, the main goal of active learning is to maximize the value of the obtained preference data at the minimum cost. This is typically done by analyzing the dataset and actively selecting a restricted set of items to ask the user to rate, hoping that 
 the new data will improve 
the most the performance of the system. Therefore, it is very important for 
the active learning system to define a precise procedure to select 
the most beneficial items, which is called {\it Strategy}.

So far,  a variety of active learning strategies have been proposed and evaluated \cite{rubens2015active,elahi2016survey}.
However, almost all previous research works have conducted their evaluations under a specific scenario, i.e., active learning in the single-domain recommendation scenario.  
This is while many real-world RSs are actually operating in
the cross-domain scenario, where the user preferences are available 
not only in the target domain, but also in the additional auxiliary domain. Knowledge of user preferences in the auxiliary domain can be \textit{transferred} to the target domain to mitigate, among others, the effect of the new user problem.
However, to the best of our knowledge, almost no work has previously focused on
active learning in cross-domain recommendation  scenario. 

In this  paper, we address this limitation by implementing a number of widely used  active learning strategies and evaluating them 
in the cross-domain recommendation scenario. These strategies 
have been already implemented and thoroughly evaluated in the single-domain scenario \cite{rubens2015active,elahi14active,elahi2014survey,elahi2012adapting,Rashid08learningpref,Golbandi11Adaptive}. We extend them to be compatible with the usage across domains and evaluate them in a novel evaluation framework explicitly suited for this scenario. 

We have performed an offline experiment  in order to
measure the performances of the considered strategies and compared them
with respect to two different evaluation metrics, i.e., prediction accuracy (in terms of MAE), and coverage (in terms of Spread). 
Our results have shown that, while active learning strategies can still effectively improve the system in both scenarios, 
the performance of these strategies can be significantly
changed in the cross-domain scenario in comparison to single-domain scenario.   
In fact, this indicates the need for effective active learning  strategies  in RSs
and still
provides a realistic and accurate guidelines applicable to real-world RSs.


\section{Related Work}
\label{sec:relwork}

\subsection{Cross-domain RSs}
The cross-domain recommendation problem has been studied from multiple perspectives in different areas. In user modelling, solutions have been proposed that aggregate and mediate user preferences \cite{abel2013,berkovsky2008,fernandez2016alleviating,cremonesi2011cross,cantador2015cross}, and in machine learning it has been studied as a practical application of knowledge transfer techniques \cite{gao2013,zhao2013active,cremonesi2014cross}. In the area of recommender systems, cross-domain recommender systems have been proposed as a solution to the cold-start and sparsity problems \cite{enrich2013,sahebi2013}. 
For instance, 
Shahebi and Brusilovsky \cite{sahebi2013} analyzed the effect of the user's profile size in the auxiliary domain and showed that, when enough auxiliary ratings are available, cross-domain collaborative filtering can provide better recommendations in the target domain in cold-start situations. 

In this paper, we aim at addressing the new user problem from a different
perspective. Rather than exploiting auxiliary information directly in the
prediction model, we are interested in improving the active learning phase of the system. Our proposed approach exploits user preferences from an auxiliary domain in order to better select the ratings to be elicited from the user in the target domain, by means of active learning. 

In this regard, \cite{zhang2016multi} is the only work that is comparable with our work.
However, that work is still different from our work mainly in the evaluation methodology
where they have assumed that the simulated users can rate 400 items in every learning iteration.
Accordingly, they iterate for 50 times and ask the the simulated users to rate 20000 items.
However, we assume that the users may rate   5 items which seems a more realistic assumption.
Moreover, they only evaluate the performance of active learning strategies in the cross-domain
scenario while we  perform the evaluation in both single-domain and cross-domain scenarios.
Finally, they only consider prediction accuracy (in terms of RMSE) as the evaluation metric.
However, while prediction accuracy is classical metric for evaluation of RSs, still
do not reflect other important aspects of the recommendation quality. Hence, 
we evaluate the performance of active learning strategies no only  in terms of prediction accuracy (MAE), 
but also Spread which is an indication of how well the recommended items are diversified
\cite{kluver2014evaluating,fernandez2016alleviating}. 
%

\subsection{Active Learning in RSs}
There are a broad range of active learning strategies that have been already 
proposed and evaluated \cite{rubens2015active,elahi14active,Rashid08learningpref,elahi2014survey,Golbandi11Adaptive,fernandez2016alleviating}.
Among these strategies, we implemented a number of widely used and best performing strategies.
In spite of excellent performance, we could not consider one type of strategies that are based on decision trees. This is due to the fact that
 their computational complexity grows directly with the number of items in the dataset, which in our tests, is considerably bigger than the ones performed in the original paper \cite{Golbandi11Adaptive}. More importantly, the number of users in our dataset is not sufficient to properly build a tree deep enough,   without incurring into overfitting.

We now describe the active learning strategies that we evaluated in our experimental study. 

\begin{description}

\item{\it Highest-predicted} \cite{elahi14active,rubens2015active}: scores items according to the rating prediction
values and selects the top items according to
their scores. The items with the highest predicted 
ratings are the ones that the system expects the user likes the most.
Hence, it could be more likely that the user have experienced these items. 

\item{\it Lowest-predicted} \cite{elahi14active,rubens2015active}: uses the opposite heuristics compared to
highest predicted: the score assigned by the strategy to each item is
 $Maxr - \hat{r}$, where $Maxr$ is the 
maximum  rating value (e.g., 5) and the $\hat{r}$ is the predicted rating. This ensures that items with 
the lowest predicted ratings will get the highest 
score and therefore will be selected for elicitation. Lowest predicted 
items are likely to reveal what the user dislikes, but are also likely to elicit a
few ratings, since users tend to not to rate items that they do not like.

\item{\it Entropy0} 
\cite{Rashid08learningpref,Golbandi11Adaptive}:
measures the dispersion of the ratings 
computed by using the relative frequency of each of the five possible rating values (1-5), and also the unknown ratings as a new rating value, equal to 0, and hence considering a rating scale between 0 to 5. 
In such a way, a high frequency of the 0 rating (i.e., many unknown ratings) tends to  decrease Entropy0. Hence, this strategy 
favors popular items that are informative at the same time
\cite{Rashid08learningpref}.

\item{\it Popularity} \cite{Rashid08learningpref,rashid2002getting}:  selects  the 
items that have received the highest number of ratings. 
Such items are more likely to be known by the user, and consequently have higher chances to be rated and thus to increase data available for the RS  \cite{Carenini2003Toward}.
\end{description} 

  \begin{figure*}[ht]
	\centering
 	\includegraphics[scale=0.45]{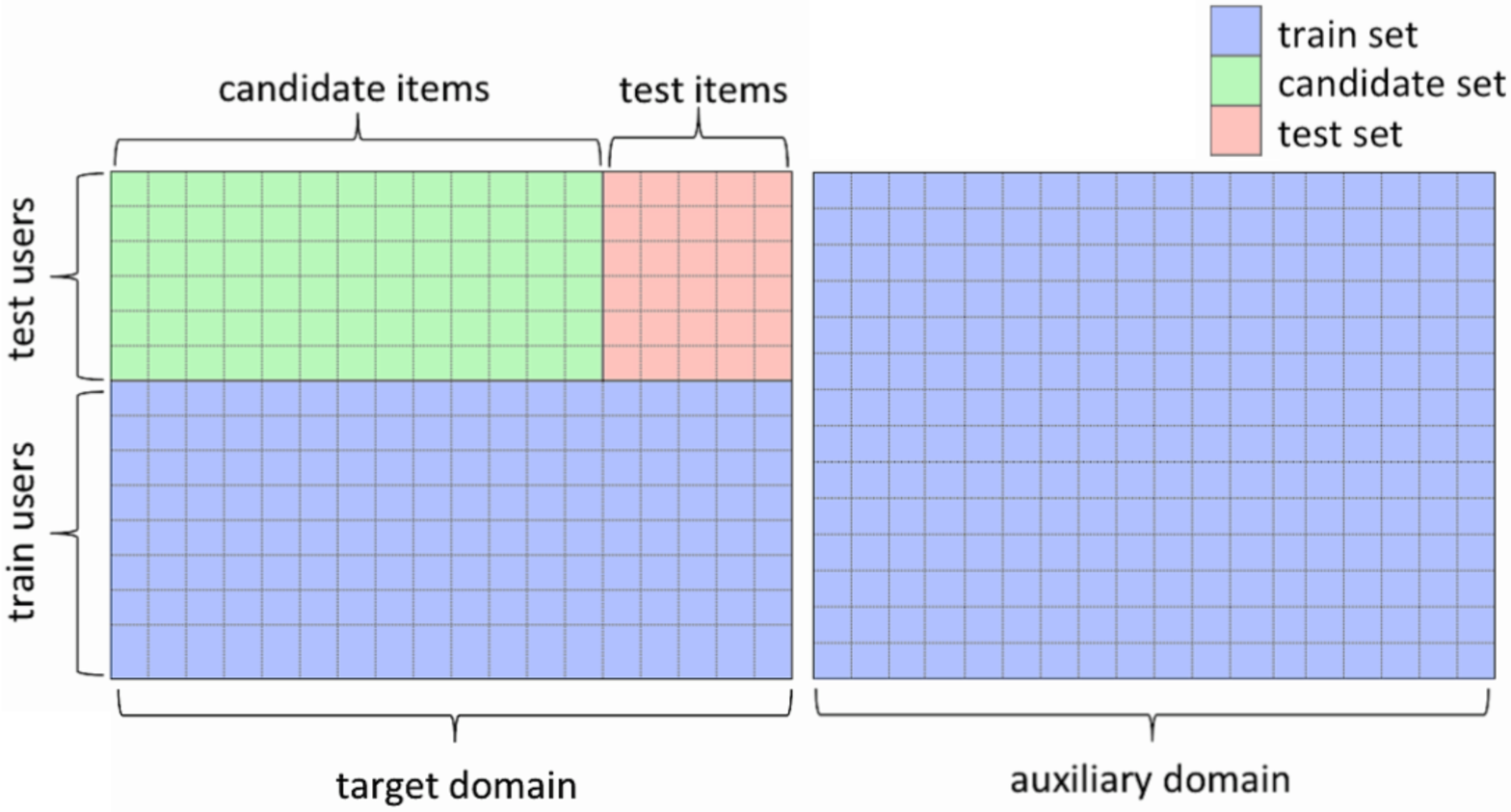}
 	\caption{Evaluation framework for new user problem in cross-domain scenario}
	\label{fig:eval-meth}
\end{figure*}

\section{Experimental Evaluation}
\label{sec:evaluation}

\subsection{Dataset}
\label{subsec:dataset}

In our experiments, we used the Amazon SNAP  \footnote{\tiny\url{https://snap.stanford.edu/data/web-Amazon.html}}, that comprises the user preferences on variety of product domains. Among these domains, we selected the two most overlapping ones, namely MoviesAndTv and Music. 
Then we considered only the common users between the two domains with 
at least 20 ratings in each domain (minimum 40 ratings in both domains). The resulting dataset has 796,489 movie ratings and 436,446 music ratings,  given by 2786 users to 86,206 movies (0.32\% density), and 151,368 tracks (0.10\% density), respectively. We note that the MoviesAndTv
dataset  is considered  as target and Music dataset as auxiliary.

\subsection{Evaluation Methodology}
\label{sec:evaluation_setting}

We evaluated the active learning strategies using an evaluation methodology similar to the one   proposed in \cite{kluver2014evaluating,fernandez2016alleviating}. It basically uses a user-based 5-fold cross-validation, that we modified 
for active learning process in cross-domain scenario. For our experiments we employed the FunkSVD matrix factorization available in the LensKit Framework~\cite{Ekstrand:2011} as recommendation algorithm.

First, we shuffle the set of users in the target domain and split it into 5 disjoint subsets of equal size. In each cross-validation step, the ratings from 4 subsets are used to train the recommendation algorithm and the active learning strategy. The ratings for the users in the remaining subset are further  split into 3 randomly generated subsets (see figure \ref{fig:eval-meth}):
{\bf Train set} contains the set of known ratings for each user. We simulated different profile sizes for each new user by incrementally adding one rating at time.
{\bf Candidate set} contains the set of ratings that can be elicited by the active learning strategy and then added to the train set (at least 15 ratings).
{\bf Test set}  contains the set of ratings used to compute the performance metrics (5 ratings per user).

In order to apply active learning to the cross-domain scenario,
we have extended the training data by adding to ratings of the target domain the entire set of ratings in the auxiliary domain. Such extended training set was provided as input to the recommendation algorithm to then generate rating predictions.

Then, we follow the evaluation procedure originally proposed in \cite{elahi14active} and described below:

\begin{enumerate}
\item We train the recommendation algorithm on the train set and then we compute the evaluation metrics on the test set.
\item For every test user, the active learning strategy ranks each item in the candidate set. The rating of the top ranked candidate item is added to the train set and removed from the candidate set.
\item We repeat the procedure with the new train and candidate sets.
\end{enumerate}

In all of the experiments, we considered profiles starting from no rating, i.e., the \textit{Extreme New User} problem, and constantly increased with one rating at time until to the limit of  5 ratings per user profile has been reached. 

We have considered two evaluation metrics, namely  Mean Average Error (MAE), i.e., 
the mean absolute deviation of the predicted ratings from the actual ratings \cite{shani10}
and  {Spread}, i.e.,  a metric of how well the recommender or active 
learner spreads its attention across many items with the assumption that better algorithms 
select  different items for different users \cite{kluver2014evaluating,fernandez2016alleviating}. 
 
 
\section{Results and Discussion}
\label{sec:result}

  \begin{table*}[ht]

\begin{center} 
    \begin{tabular}{   c | l | c | c | c | c | c | c | c | c |   }
\cline{2-10}
&   {\multirow{3}{*}{\bf AL Strategy} }       & \multicolumn{4}{c |}{\bf MAE} &    \multicolumn{4}{c |}{\bf Spread }    \\
  &     &  \multicolumn{2}{c |}{\bf Single-domain}            &  \multicolumn{2}{c |}{\bf Cross-domain}          & \multicolumn{2}{c |}{\bf  Single-domain}                 &    \multicolumn{2}{c |}{\bf Cross-domain}          \\
            \cline{3-10}
  &    &  value & improve            & value & improve         & value & improve            &   value & improve         \\
            \hline

\multicolumn{1}{|l|}{\multirow{4}{*}{\bf with AL}}   & High-predicted   \cite{elahi14active,rubens2015active}   & {\bf 0.823}  & {\bf 8.6\%} & 0.811         & <1.0\% & 3.352  & 111.4\% & 6.533  &  2.9\%     \\
\cline{2-10}
\multicolumn{1}{|l|}{}&  Low-predicted   \cite{elahi14active,rubens2015active}    & 0.837  & 7.1\% & \textbf{0.807 }      & {\bf 1.1\%} & {\bf 5.063} & {\bf 219.4\%}  & 6.958    &   9.6\%   \\
\cline{2-10}
\multicolumn{1}{|l|}{}&  Popularity \cite{Rashid08learningpref,rashid2002getting}     & 0.826  & 8.3\% & 0.811        & <1.0\% & 4.693  & 196.0\% & \textbf{6.968  }   &  {\bf 9.8\%}   \\
\cline{2-10}
\multicolumn{1}{|l|}{} &   Entropy0 \cite{Rashid08learningpref,Golbandi11Adaptive}    &   0.826  & 8.3\% & 0.810    &  <1.0\% & 4.704  & 196.7\% & 6.956   &  9.6\%  \\
\hline
 \multicolumn{1}{|l|}{\bf without AL}  & -   &    0.901  & - & 0.816     &  - & 1.585  & - & 6.346   &  -  \\
\hline
    \end{tabular}
    \end{center}

     \caption{The  performance of the active learning strategies in two recommendation scenarios: (i) single-domain scenario (without any auxiliary domain), and (ii)  cross-domain scenario (with auxiliary domain)}
    \label{tab:comparison-aux}
\end{table*}

Table~ \ref{tab:comparison-aux} shows the performance of different active learning strategies, 
in single-domain and cross-domain recommendation scenarios. 
The performance comparison has been made with respect to MAE and Spread metrics
and the improvement of the RS {\bf with AL} over the RS {\bf without AL} in both scenarios.

First of all, it is clear that all the active learning strategies have improved the quality of the
recommendation in either of the single-domain or the cross-domain scenarios. 
However, the improvement made by each of these strategies differs very much in these two scenarios. 
While in single-domain scenario, the best strategy is highest-predicted with MAE of 0.823 (8.6\% of improvement), 
in cross-domain scenario, the lowest-predicted strategy outperforms 
the other strategies by achieving the MAE of 0.807 (1.1\% of improvement). 
In terms of Spread, in single-domain scenario, lowest-predicted is the best strategy 
with Spread value of 5.063 (219.4\% improvement), while in cross-domain scenario,
the popularity strategy achieves the best result which is 6.968 (9.8\% improvement).
These are promising results since active learning strategies can achieve such improvements 
by only eliciting 5 ratings per user in the target domain.

%
%

We have also observed that, the initial MAE in the cross-domain scenario is much lower in comparison to single-domain scenario. 
Similarly, the Spread values in cross-domain is much higher than the values in single-domain 
scenario. This confirms that the exploitation of the ratings in the auxiliary domain significantly helps the RS to improve its 
performance in terms of these metrics.

Accordingly, while model-based active learning strategies, that are typically personalized strategies (such as lowest-predicted), may exploit 
such data to make better item selection for the users, the other strategies, that are typically non-personalized (such as popularity) cannot exploit the additional knowledge provided by the auxiliary domain. This is a big limitation of second type of active learning strategies in cross-domain scenario.
 However, their performance can still impact the quality of the data fed to the RS,
 and hence, affect the output of the system. 

\section{Conclusion and Future Work}
\label{sec:conclusion}
In this paper, we have evaluated several widely used active learning strategies adopted
to tackle the cold-start problem in a novel usage scenario, i.e., Cross-domain recommendation scenario. In such a case, the user preferences are available 
not only in the target domain, but also in additional auxiliary domain. Hence, the active learner can exploit such knowledge to better estimate which preferences 
are more valuable for the system to acquire. 

Our results have shown that the performance of the considered active learning strategies 
significantly change in the cross-domain recommendation scenario in comparison to the single-domain recommendation. Hence,
the presence of the auxiliary domain may strongly influence  the performance 
of the active learning strategies. Indeed,  while a  certain active learning strategy  
performs the best for MAE reduction in the single scenario (i.e., highest-predicted strategy), 
it   actually performs poor in the cross-domain scenario. On the other hand,
the strategy with the worst MAE in single-domain scenario (i.e., lowest-predicted strategy) 
can perform excellent  in the cross-domain scenario. This is an interesting 
observation which indicates the importance of further analysis of these two scenarios
in order to better design and develop active learning strategies for them.  
 



  
Our future work includes the further analysis of the 
AL strategies in other domains such as book, electronic products, tourism, etc.
Moreover, we plan to investigate the potential impact of considering different rating 
prediction models (e.g., context-aware models) on the performance of different active learning strategies.


\bibliographystyle{abbrv}
\bibliography{arxiv2017}
\end{document}